\documentclass{llncs}
\usepackage{llncsdoc}
\usepackage{graphicx}
\begin{document}

\title{An OpenSHMEM Implementation for the Adapteva Epiphany Coprocessor}

\author{
	James Ross\inst{1}
	\and
	David Richie\inst{2}
}

\institute{
	U.S. Army Research Laboratory, Aberdeen Proving Ground, MD 21005, USA
	\email{james.a.ross176.civ@mail.mil}
	\and
	Brown Deer Technology, Forest Hill, MD 21050, USA
	\email{drichie@browndeertechnology.com}
}

\maketitle

\begin{abstract}

This paper reports the implementation and performance evaluation of the
\mbox{OpenSHMEM~1.3} specification for the Adapteva Epiphany architecture
within the Parallella single-board computer. The Epiphany architecture exhibits
massive many-core scalability with a physically compact 2D array of RISC CPU
cores and a fast network-on-chip (NoC). While fully capable of MPMD execution,
the physical topology and memory-mapped capabilities of the core and network
translate well to Partitioned Global Address Space (PGAS) programming models
and SPMD execution with SHMEM.


\end{abstract}

\section{Introduction and Motivation}

The \mbox{OpenSHMEM} communications library is designed for computer platforms
using Partitioned Global Address Space (PGAS) programming models
\cite{Chapman}. Historically, these were large Cray supercomputers, but now the
\mbox{OpenSHMEM} interface may also be used on commodity clusters. The Adapteva
Epiphany architecture represents a divergence in computer architectures
typically used with \mbox{OpenSHMEM} and is just one of many emerging parallel
architectures that present a challenge in identifying effective programming
models to exploit them. While some researchers may be considering how the
\mbox{OpenSHMEM} API may interact with coprocessors, the work presented here
leverages the API for device-level operation. In some aspects, the Epiphany
architecture resembles a symmetric multiprocessing (SMP) multi-core processor
with a shared off-chip global memory pool. However, each core can directly
address the private address space of neighboring cores across an on-chip 2D
mesh network. Thus, the architecture also has the characteristics of a PGAS
platform. Previous proof-of-concepts demonstrated that message passing
protocols could achieve good application performance on the Epiphany
architecture \cite{Richie2015}, \cite{Implementing2016}. However, it was
unclear if the \mbox{OpenSHMEM~1.3} standard could be fully implemented within
the platform limitations and achieve high performance using a standard
programming model without resorting to non-standard software extensions.

Existing open source \mbox{OpenSHMEM} implementations are inadequate within the
constraints of the Epiphany architecture, so a new C language implementation
named \emph{ARL \mbox{OpenSHMEM} for Epiphany} was developed from scratch. The
design emphasizes a reduced memory footprint, high performance, and simplicity,
which are often competing goals. This paper discusses the Epiphany architecture
in Sect. \ref{ssec:epiphany}, the \mbox{OpenSHMEM} implementation and
performance evaluation in Sect. \ref{sec:implementation}, and a discussion of
future work and potential standard extensions for embedded architectures in
Sect. \ref{sec:future}.

\section{Background}
\label{sec:background}

The 16-core Epiphany-III coprocessor is included within the \$99 ARM-based
single-board computer and perhaps represents the low-cost end of programmable
hardware suitable for SHMEM research and education. Many universities,
students, and researchers have purchased the platform with over 10,000 sales to
date. Despite this, programming the platform and achieving high performance or
efficiency remain challenging for many users. Like GPUs, the Xeon Phi, and
other coprocessors, typical applications comprise host code and device code.
Only a minimal set of communication primitives exist within the non-standard
Epiphany Hardware Utility Library (eLib) for multi-core barriers, locks, and
data transfers \cite{esdkgithub}. The barrier and data transfer routines are
not optimized for low latency. Other primitives within eLib use unconventional
2D row and column indexing, which cannot easily address arbitrary numbers of
working cores or disabled cores. More complicated collectives, such as those in
the \mbox{OpenSHMEM} specification, are left as an exercise for the application
developer.

Although not discussed in detail in this paper, the CO-PRocessing Threads
(COPRTHR) 2.0 SDK \cite{COPRTHR2} further simplifies the execution model to the
point where the host code is significantly simplified, supplemental, and even
not required depending on the use case \cite{Advances2016}. There are
essentially two modes of possible execution. The first mode requires host code
with explicit Epiphany coprocessor offload routines. The second mode uses a
host-executable coprocessor program with the conventional main routine
provided. The program automatically performs the coprocessor offload without
host code. Combined with the work presented in this paper, the
\mbox{COPRTHR~2.0} SDK enables many \mbox{OpenSHMEM} applications to execute on
the Epiphany coprocessor without any source code changes.  Execution occurs as
if the Epiphany coprocessor is the main processor driving computation.
COPRTHR~1.6 was used to present the Threaded MPI model for Epiphany
\cite{Richie2015} as well as a number of applications \cite{HPEC2015},
\cite{Ross2016}.

\subsection{Epiphany Architecture}
\label{ssec:epiphany}

\begin{figure}
	\centering
		\includegraphics[width=1.0\textwidth]{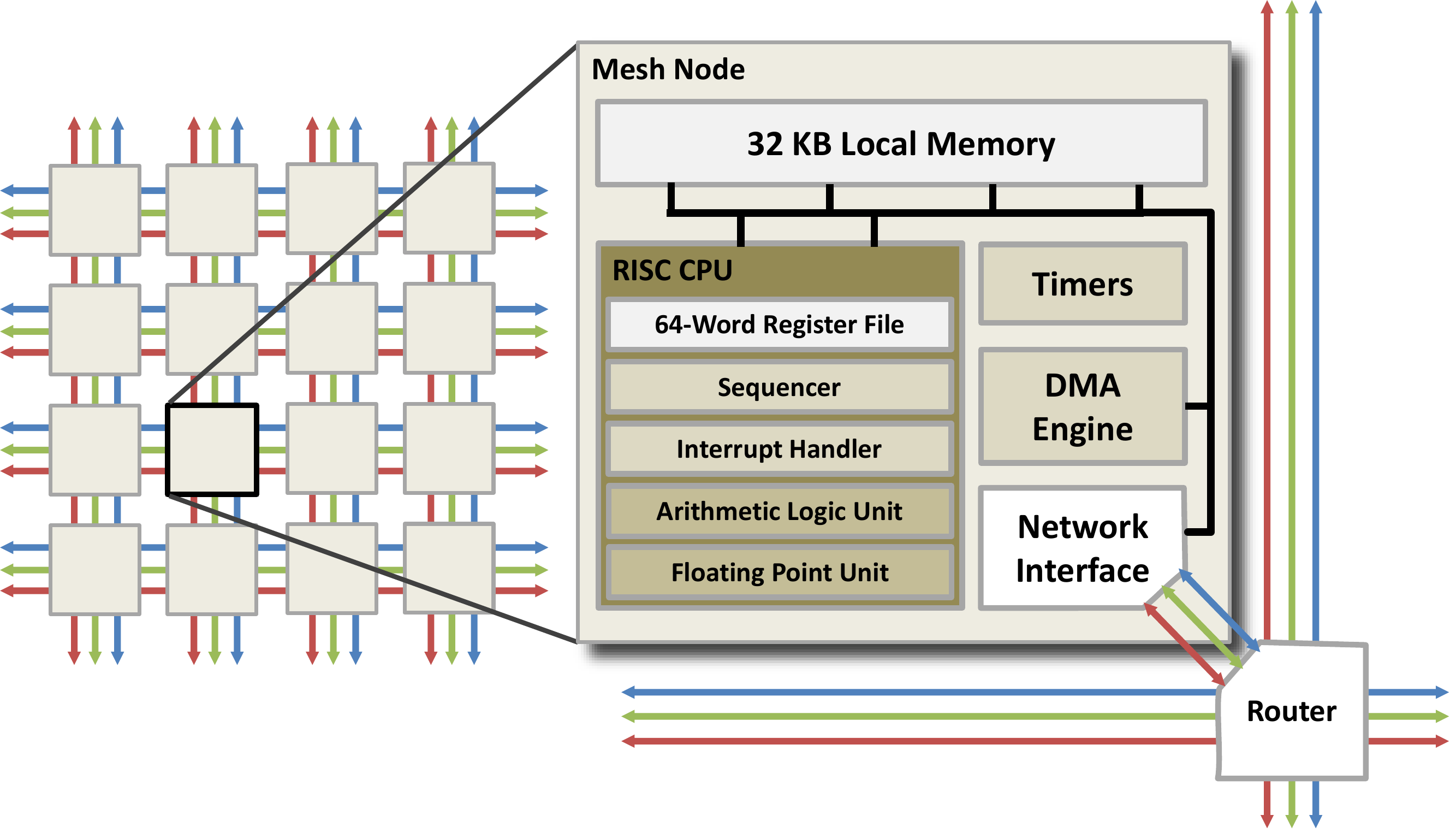}
	\caption{The 16-core Epiphany-III architecture is a 2D array of RISC CPU
cores. It contains a 64-word register file, sequencer, interrupt handler,
integer and floating point units, timers, and DMA engines for the fast
network-on-chip}
	\label{fig:epiphany}
\end{figure}

Many modern computer architectures address the ``memory wall problem'' by
including increasingly complex cache hierarchies and core complexity, wider
memory buses, memory stacking, and complex packaging to maintain the SMP
hardware and software architecture. The Epiphany architecture unwinds decades
of these types of changes -- it is a cache-less, 2D array of RISC cores with a
fast network-on-chip (NoC) that can an be simply described as a ``cluster on a
chip''. Each core within the Epiphany-III architecture contains 32 KB of SRAM
which is shared between instructions and local data. The Epiphany architecture
can scale to one megabyte of SRAM per core, but there is a linear design
tradeoff between the number of cores and available memory for a fixed die
space. The core local memory is memory-mapped, and each core may directly
access the local memory of any core within the mesh network. Each core has
shared memory access to off-chip global DRAM, although this access is
significantly slower than local memory or non-uniform memory access (NUMA) to
neighboring core memory. The highest performance and most energy-efficient
applications leverage inter-core communication and on-chip data reuse. Like
many high performance computing (HPC) clusters, the inter-core communication is
generally explicit in order to achieve highest performance. The architecture is
also scalable by tiling multiple chips without additional ``glue logic''. The
tight coupling between the core logic and the on-chip mesh network enables very
low-latency operation of \mbox{OpenSHMEM} routines. An architectural overview
appears in Figure~\ref{fig:epiphany}. Unlike most application programming
interfaces for communication, there is no additional software layer to handle
networking for hardware abstraction. As we will discuss in further detail, the
\mbox{OpenSHMEM} implementation for Epiphany performs network operations
directly.

\section{Implementation and Performance Evaluation}
\label{sec:implementation}

Due to the tight memory constraints of the Epiphany memory and availability of
specialized hardware features, the \mbox{OpenSHMEM} reference implementation
built on GASNet was not suitable for deployment on the Epiphany cores. As a
credit to the \mbox{OpenSHMEM} specification and the Adapteva Epiphany
architecture documentation, the full \mbox{OpenSHMEM~1.3} implementation was
written and optimized over a period of a few weeks. The entire library,
including the optional extensions described in detail later, is approximately
1800 lines of code and does not require additional software. The software
directly targets the underlying hardware features and was designed to be
extremely lightweight in order to compile to small binaries expected with
embedded architectures.

Linear scaling algorithms were avoided, and many of the collective routines use
dissemination or recursive doubling algorithms, optimized for low-latency on
the Epiphany network. The remote memory access routines,
\texttt{shmem\_\emph{TYPE}\_put} and \texttt{shmem\_\emph{TYPE}\_get}, use
hand-tuned memory-mapped load and store primitives with a hardware loop feature
specific to the Epiphany architecture. The non-blocking remote memory access
routines use the dual-channel Direct Memory Access (DMA) engine on each
processor network node. The distributed locking and atomic routines leverage an
atomic \texttt{TESTSET} instruction that performs an atomic
``test-if-not-zero'' and conditional write. An optional hardware barrier
implementation was also developed for a specialized
\texttt{shmem\_barrier\_all} for extremely low-latency global barriers. An
optional inter-processor interrupt and corresponding interrupt service routine
(ISR) enable faster \texttt{shmem\_\emph{TYPE}\_get} operations by interrupting
the remote core to use the optimized \texttt{shmem\_\emph{TYPE}\_put}.

Many of the \mbox{OpenSHMEM} routines have some component that is hardware
accelerated on the Epiphany architecture such as zero-overhead hardware loops
for copying data, memory-mapped loads and stores, the \texttt{TESTSET}
instruction for remote locks and atomics, a wait on AND (\texttt{WAND})
instruction for a low-latency \texttt{shmem\_barrier\_all}. The
\texttt{MULTICAST} experimental feature would enable energy-efficient,
low-latency broadcasts but is presently unused. The point-to-point
synchronization routines are among the simplest to implement and do not have a
section dedicated to discussion. Generally, they spin-wait on local values
until they meet the criteria defined by the routine. The memory ordering
routines need only verify that both DMA engines have an idle status by
spin-waiting on the relevant special register. There are no intermediate data
copies in this implementation.

The performance evaluation of the Epiphany \mbox{OpenSHMEM} implementation
began with the \mbox{OpenSHMEM} micro-benchmark codes. The timing code had to
be modified because the \texttt{gettimeofday} routine is only accurate to a
microsecond, and many of the operations operate in the sub-microsecond regime.

Many of the communication routines in the performance evaluation include the
parameters \(\alpha\) and \(\beta^{-1}\) in the figure subtitle along with
their standard deviations. These two parameters are from the
``\(\alpha\)-\(\beta\) model'' for communication in HPC. They neatly summarize
the communication time (\(T_c\)) to include the latency (\(\alpha\)) and
marginal cost (\(\beta\)) to transfer a message (of size \(L\)) in
equation~\ref{eq:comm}. The \(\beta^{-1}\) parameter is the peak effective core
bandwidth for the routine.

\begin{equation}
	T_c = \alpha + \beta \cdot L
	\label{eq:comm}
\end{equation}

\subsection{Library Setup, Exit, Query Routines}

The \texttt{shmem\_init} routine retrieves or calculates the local processing
element (PE) number (for \texttt{shmem\_my\_pe}) and number of PEs (for
\texttt{shmem\_n\_pes}), configures the optimized hardware barrier or
collective dissemination barrier arrays, obtains the SHMEM heap memory offset,
and precalculates a few other addresses for improved runtime performance. The
\texttt{shmem\_ptr} routine can directly calculate remote memory locations
using simple logical shift and bitwise operations.

\subsection{Memory Management Routines}
\label{ssec:memory}

\begin{figure}
	\centering
		\includegraphics[width=1.0\textwidth]{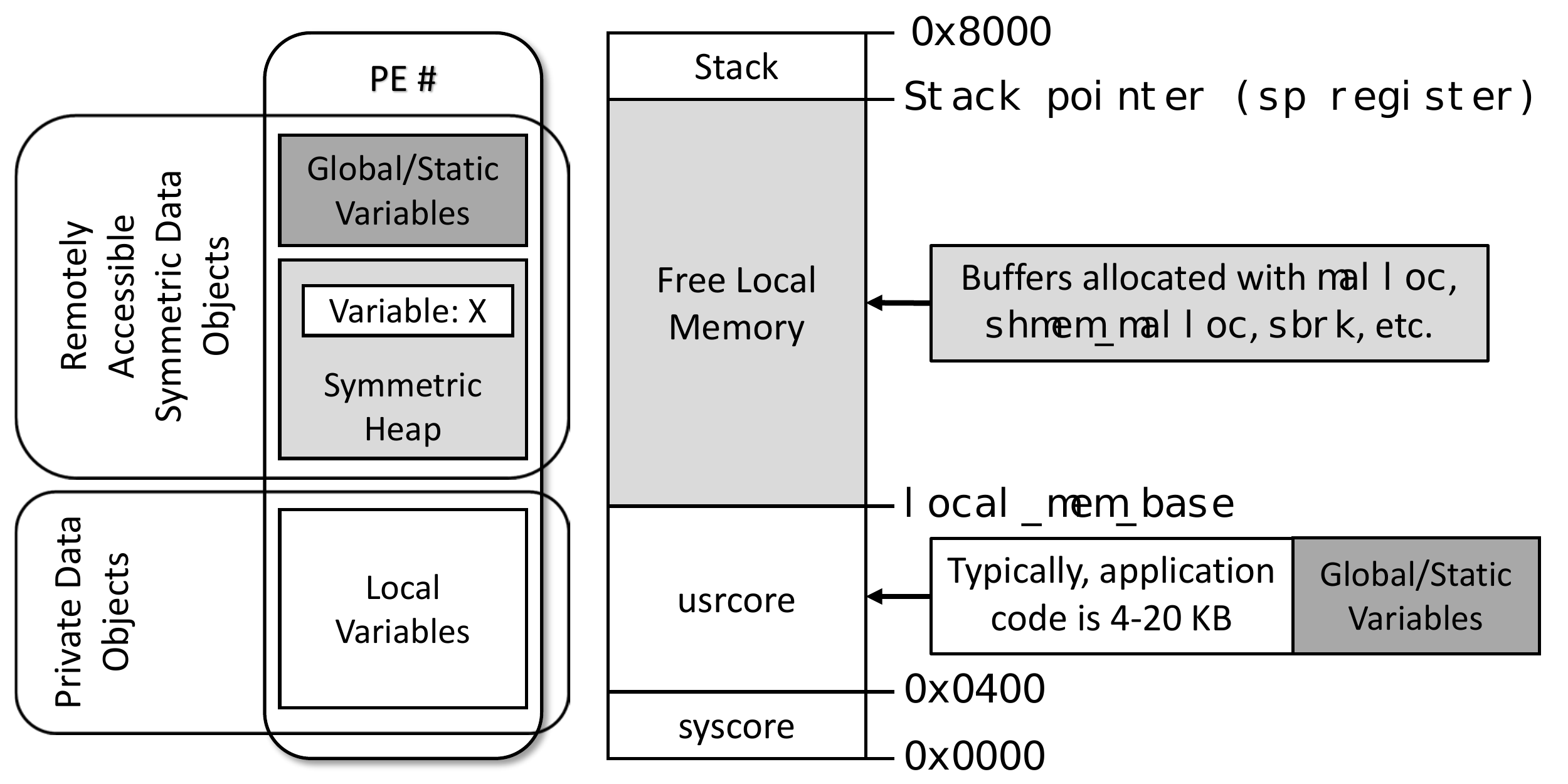}
	\caption{The PGAS memory model (left) and the equivalent typical memory
layout on an Epiphany-III core (right)}
	\label{fig:memory}
\end{figure}

Memory management on the Epiphany processor is atypical. Each Epiphany-III core
has a flat 32~KB local memory map from address \texttt{0x0000} to
\texttt{0x7fff}. Programs are typically loaded starting at \texttt{0x0100} if
extremely constrained for memory, or \texttt{0x0400} if using the COPRTHR~2
interface. The stack pointer typically moves downward from the high address.
Data used for the application, including the SHMEM data heap, begins directly
after the program space. Figure~\ref{fig:memory} shows the typical memory
layout of an Epiphany-III core using the COPRTHR~2 interface as it relates to
the PGAS model. The static or global variables that are typically defined
within the application appear below the free local memory address within the
symmetric heap. They are still symmetrical across all Epiphany cores as the
program binary is identical.

Due to the tight memory constraints, a more modern memory allocator was not
addressed in this work. The basic memory management system calls \texttt{brk}
and \texttt{sbrk} are more suited for controlling the amount of memory
allocated from the SHMEM data heap for each process element because there is no
virtual address abstraction. Instead, there is a local base memory tracking
pointer that stores the current free memory base address and incremented with
each allocation. The memory management routines build on these calls, but care
must be taken to adhere to the following rules:

\begin{enumerate}
	\item \texttt{shmem\_free} must be called in the reverse order of allocation
if making subsequent allocations
	\item \texttt{shmem\_realloc} can only be used on the last (re)allocated
pointer
	\item \texttt{shmem\_align} alignment must be a power of 2 greater than 8
(default is 8)
\end{enumerate}

This is a pragmatic approach that we feel is reasonable and won't even be
noticed on most codes. Calling \texttt{shmem\_free} moves the local base memory
tracking pointer to the address in the function argument so most routines only
need to call it once for the first allocated buffer in a series if freeing all
memory. The \texttt{shmem\_realloc} routine could be designed to copy the
contents of the old buffer to the new buffer, however, this would waste the
memory space in the original allocation (a precious commodity on the Epiphany
architecture). Future developments with COPRTHR~2 may address these
deficiencies by exporting the COPRTHR host-side memory management to the
coprocessor threads.

\subsection{Remote Memory Access Routines}

\begin{figure}
	\centering
		\includegraphics[width=0.49\textwidth]{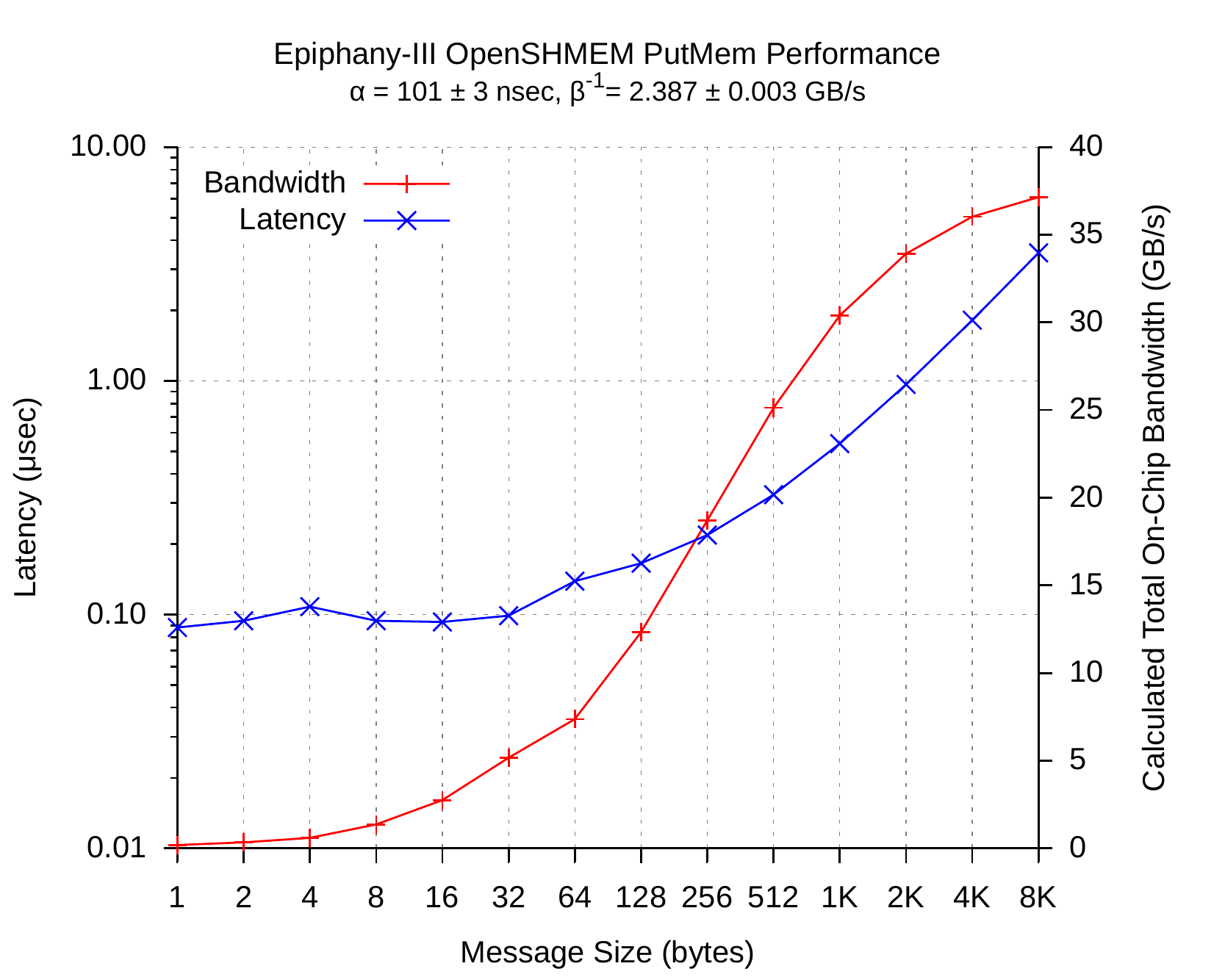}
		\includegraphics[width=0.49\textwidth]{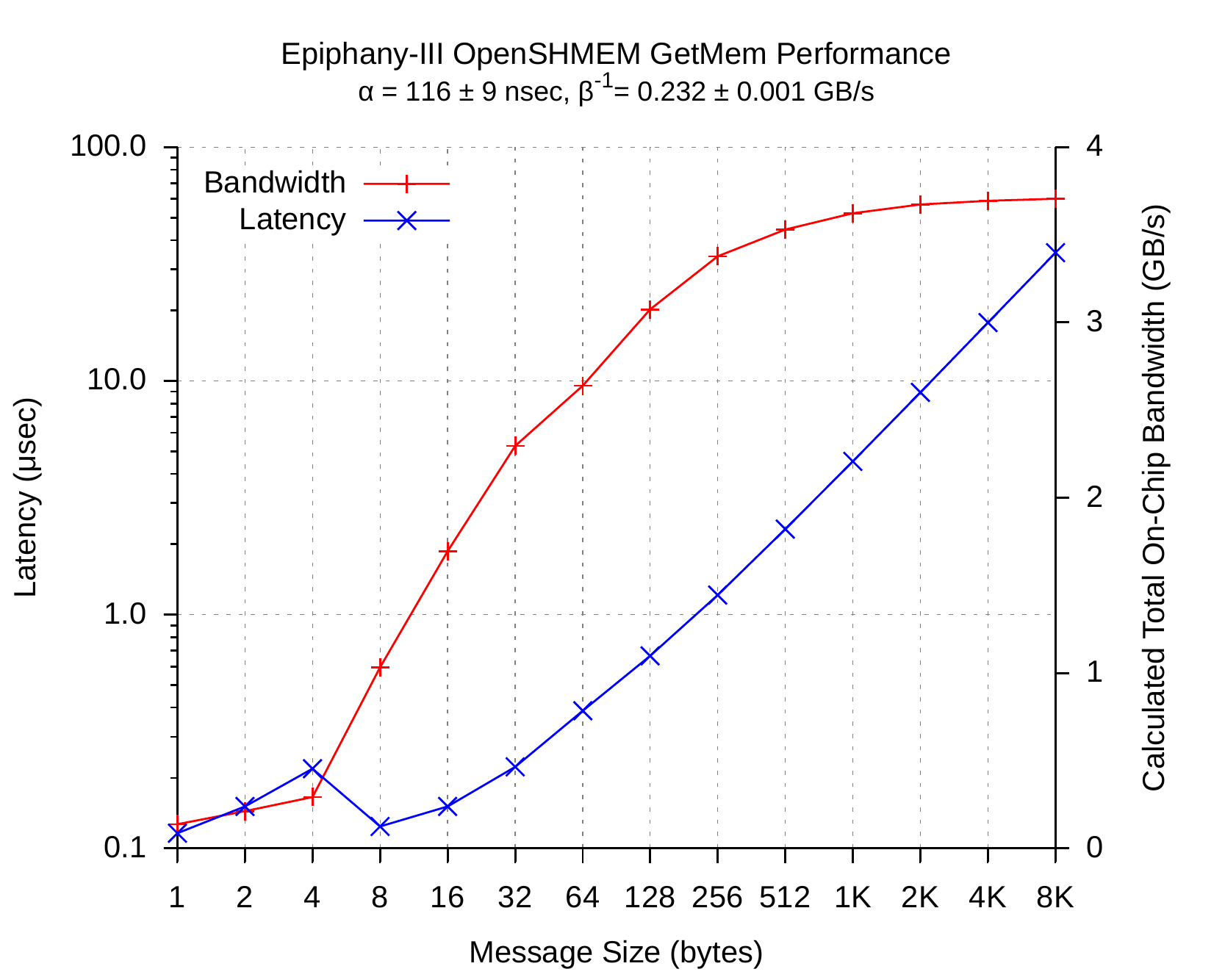}
		\includegraphics[width=0.49\textwidth]{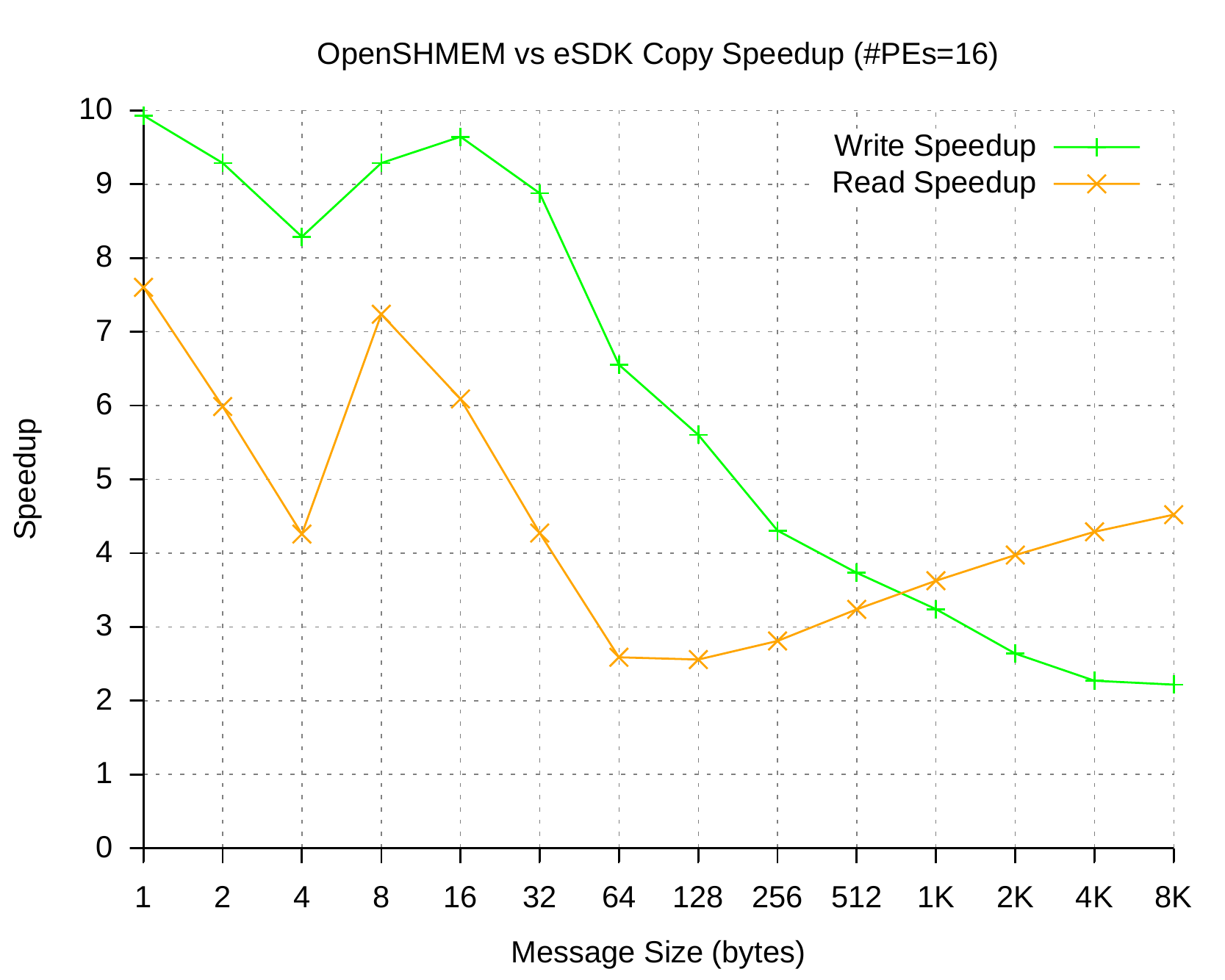}
		\includegraphics[width=0.49\textwidth]{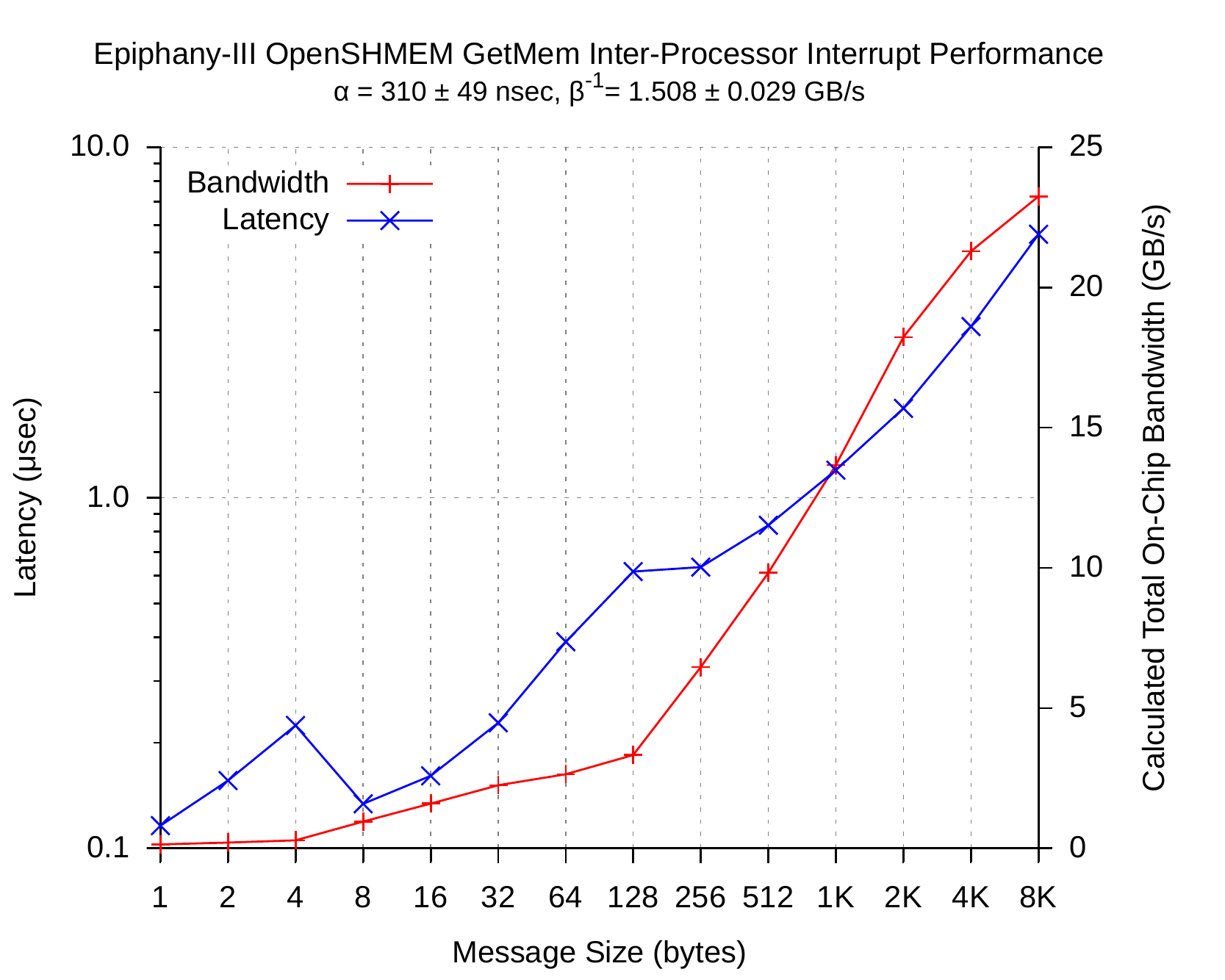}
	\caption{Performance of optimized \texttt{shmem\_put} (top left) and
\texttt{shmem\_get} (top right) for contiguous data exchange operations for 16
processing elements, speedup comparison with eLib (bottom left), and
experimental inter-processor user interrupt for high-performance
\texttt{shmem\_get} (bottom right).}
	\label{fig:put_get}
\end{figure}

Inter-process memory copying on the Epiphany is trivial, and a simple loop over
incrementing source and destination arrays can be done in C code. However, like
many optimized \texttt{memcpy} routines, high-performance copies are
non-trivial. A high-performance inter-processor memory copy routine does not
appear to be in the eLib library. So after quite some time of hand-tuning in
assembly, a put-optimized method was written that makes use of a
``zero-overhead'' hardware loop and four-way unrolled staggered double-word
loads and (remote) stores. A specialization for the edge case of unaligned
memory is also included since the Epiphany architecture requires loads and
stores to be memory aligned to the data size. Assuming the fast path is taken,
the core can transfer a double-word (8~bytes) per clock cycle. However, since
the 8~byte load operation requires an additional cycle, the effective peak
network copy is 8~bytes every two clocks. For a clock rate of 600~MHz, peak
contiguous network transfers may achieve up to 2.4~GB/sec. Having the NoC and
core clocks pinned ensures that application communication performance scales
with the chip clock speed. The same put-optimized memory copy subroutine is
used for get operations. This is sub-optimal, but remote read operations will
never be as high-performance as remote write operations on the Epiphany
architecture, so they should generally be avoided. Remote direct read
operations are slower than equivalent remote direct write operations because
the read request must first traverse the network to the receiving core network
interface, then the data must traverse the network back to the requesting core.
Unlike a remote direct write operation which can issue store instructions
without a response, the read operation stalls the requesting core until the
load instruction returns data to a register. Issuing multiple requests does
little to mitigate this performance issue, thus, the throughput of the
optimized \texttt{shmem\_put} is approximately an order of magnitude greater
than \texttt{shmem\_get} as shown in Figure~\ref{fig:put_get}.

In order to address this performance disparity with contiguous remote reads, an
inter-processor interrupt is configured and signaled by the receiving core,
causing an equivalent fast write to be executed. The receiving core is then
signaled to continue upon completion of the inter-processor ISR. This is an
experimental feature because it uses the user interrupt and must be enabled
with \texttt{SHMEM\_USE\_IPI\_GET} during compilation. It has the greatest
performance impact for large transfers. The method has a turnover point for
buffers larger than 64 bytes so that smaller transfers are read directly and
larger transfers use the inter-processor interrupt. All results for contiguous
block transfers and a performance comparison with the equivalent eLib interface
in the eSDK are shown in Figure~\ref{fig:put_get}.

\subsection{Non-blocking Remote Memory Access Routines}
\label{ssec:nonblocking}

\begin{figure}
	\centering
		\includegraphics[width=0.49\textwidth]{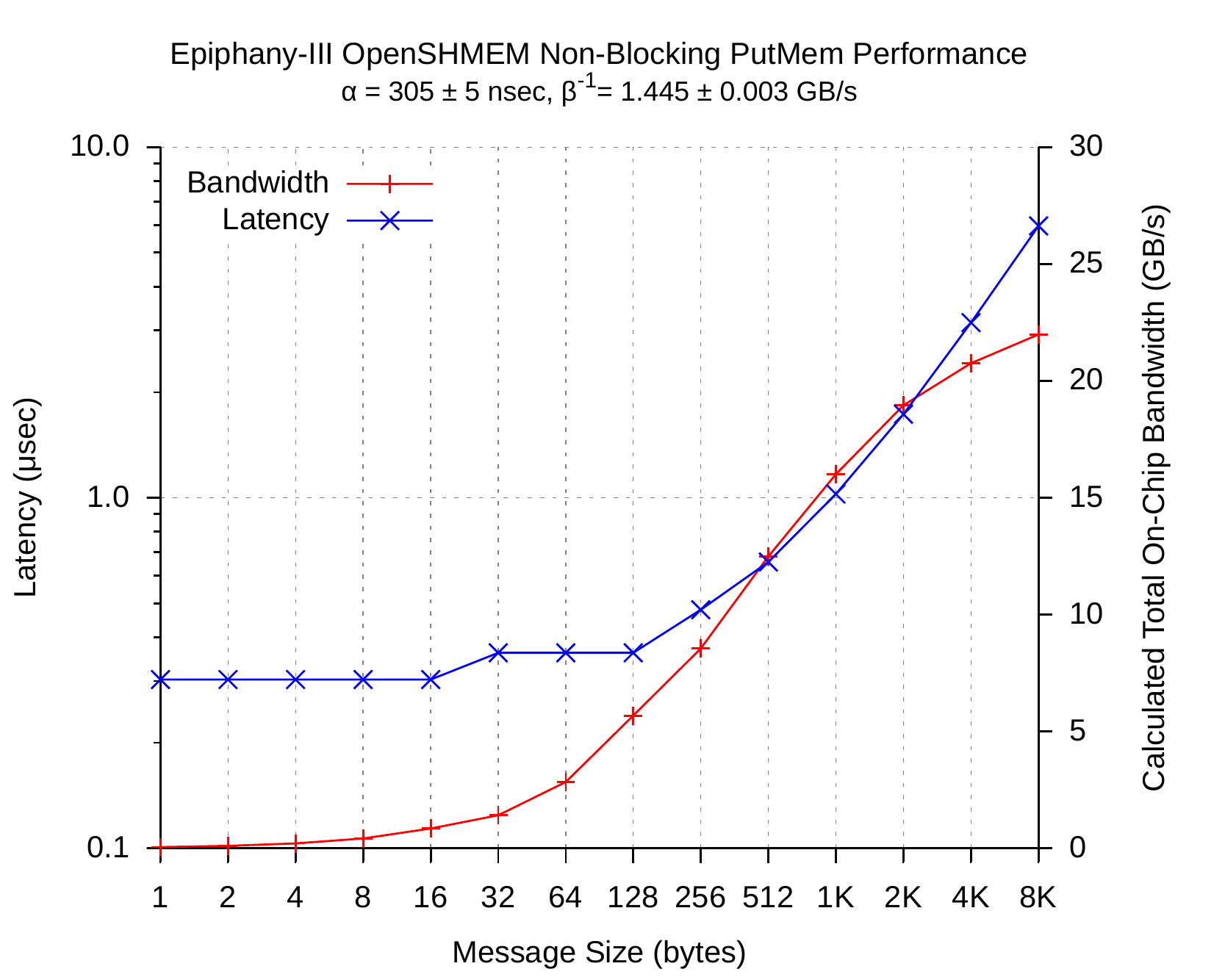}
		\includegraphics[width=0.49\textwidth]{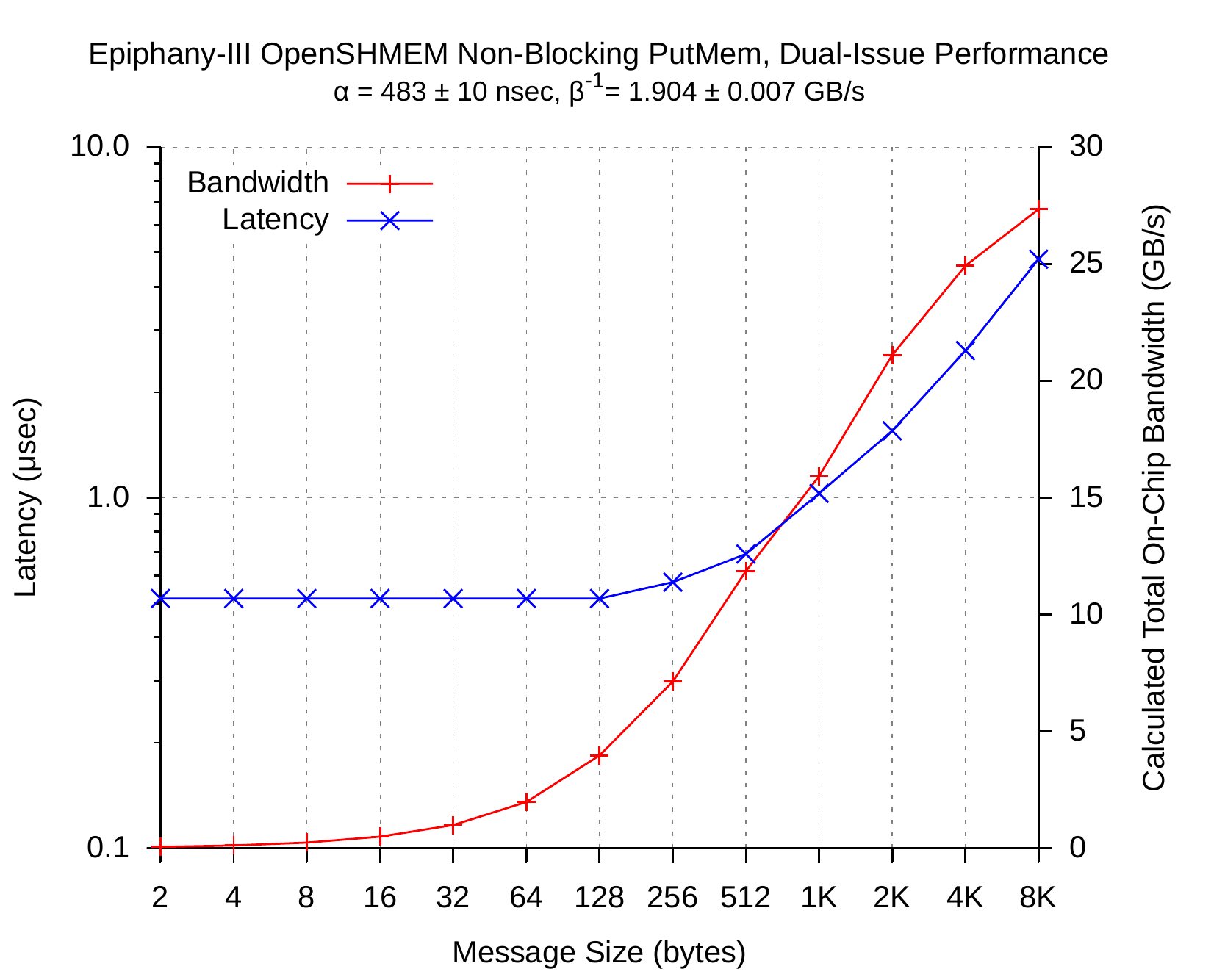}
		\includegraphics[width=0.49\textwidth]{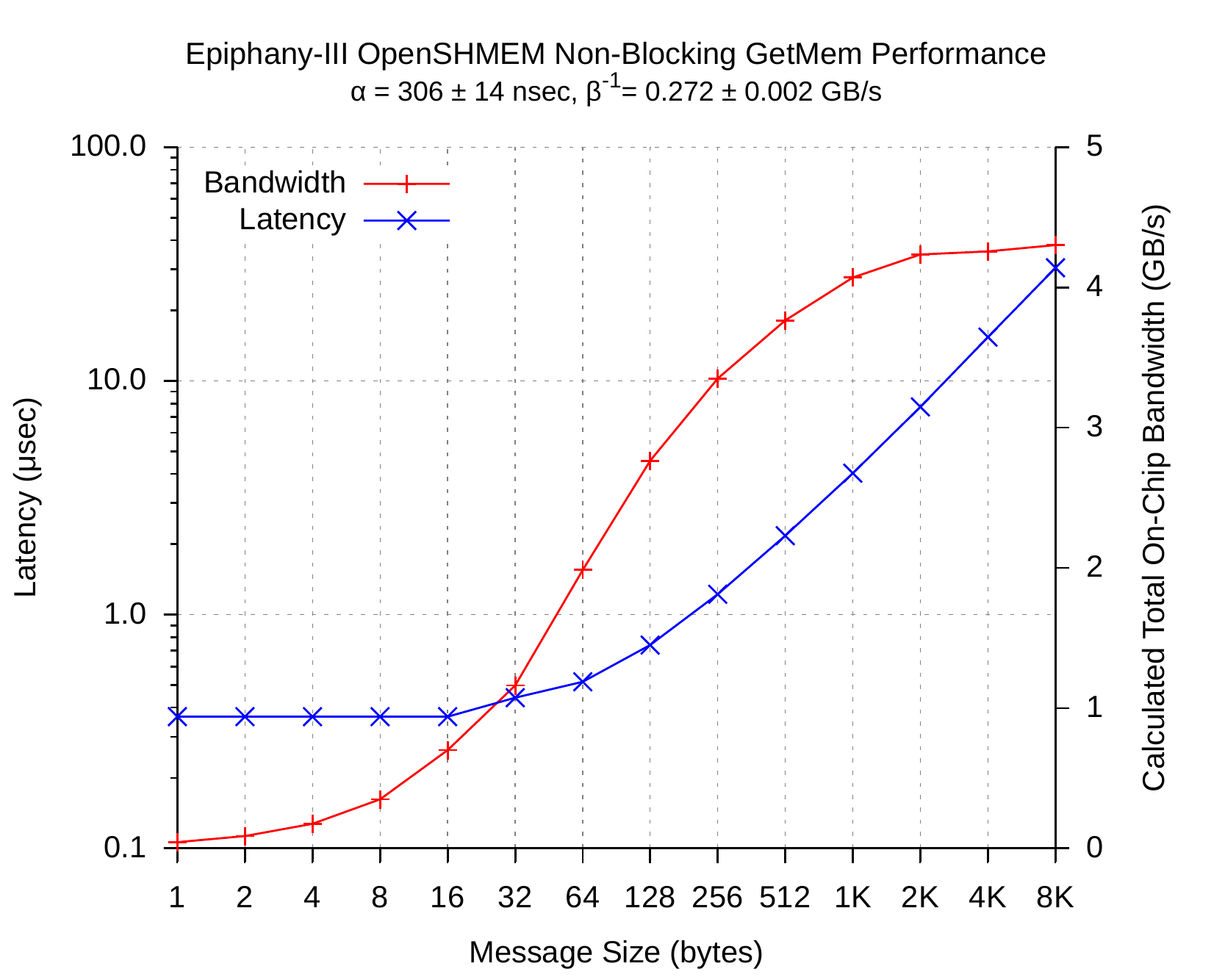}
		\includegraphics[width=0.49\textwidth]{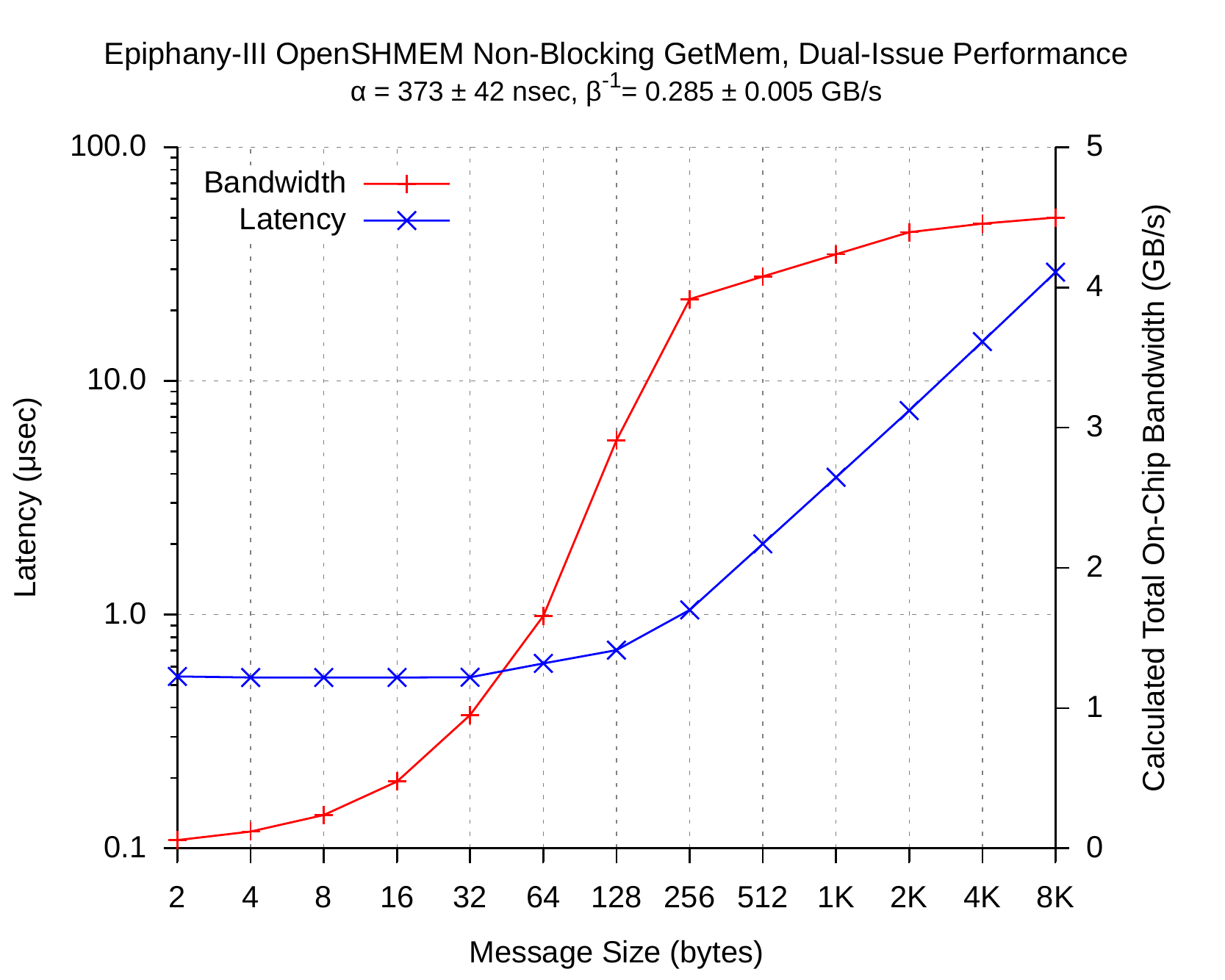}
	\caption{Performance of non-blocking remote memory access routines.}

	\label{fig:put_get_nb}
\end{figure}

The set of non-blocking remote memory access routines (\texttt{shmem\_put\_nbi}
and \texttt{shmem\_get\_nbi}) makes use of the on-chip DMA engine. The DMA
engine has two independent DMA channels per processor node so that two
non-blocking transfers may execute concurrently. Each channel has a separate
DMA specification of the source and destination address configuration. The
configuration is capable of 2D DMA operations with flexible stride sizes. This
could support an extension to the \mbox{OpenSHMEM~1.3} standard for
non-blocking strided remote memory access routines if needed. The performance
results for the non-blocking remote memory access routines appear in
Figure~\ref{fig:put_get_nb}.

Application performance improvement may be realized for large non-blocking
transfers by splitting transfers into two portions and calling two non-blocking
transfers, however, the performance benefit is marginal and often worse. Due to
hardware errata in the Epiphany-III, the DMA engine is throttled to less than
half of the peak bandwidth of 8 bytes per clock, or 4.8~GB/sec
\cite{E16datasheet}. If fully enabled, as expected in future chips, the DMA
engine would be used for the blocking remote memory access routines rather than
remote load/store instructions. In general, it may be faster to use blocking
transfers because the DMA engine setup overhead is relatively high, and there
are often bank conflicts with the concurrent computation and DMA engine access,
hindering fully overlapped communication and computation. The blocking
operation, \texttt{shmem\_quiet}, spin-waits on the DMA status register.
Alternatively, a DMA ISR could be used to continue the \texttt{shmem\_quiet}
operation, but it is not clear how this could be higher performance.

\subsection{Atomic Memory Operations}
\label{ssec:atomic}

\begin{figure}
	\centering
		\includegraphics[width=0.64\textwidth]{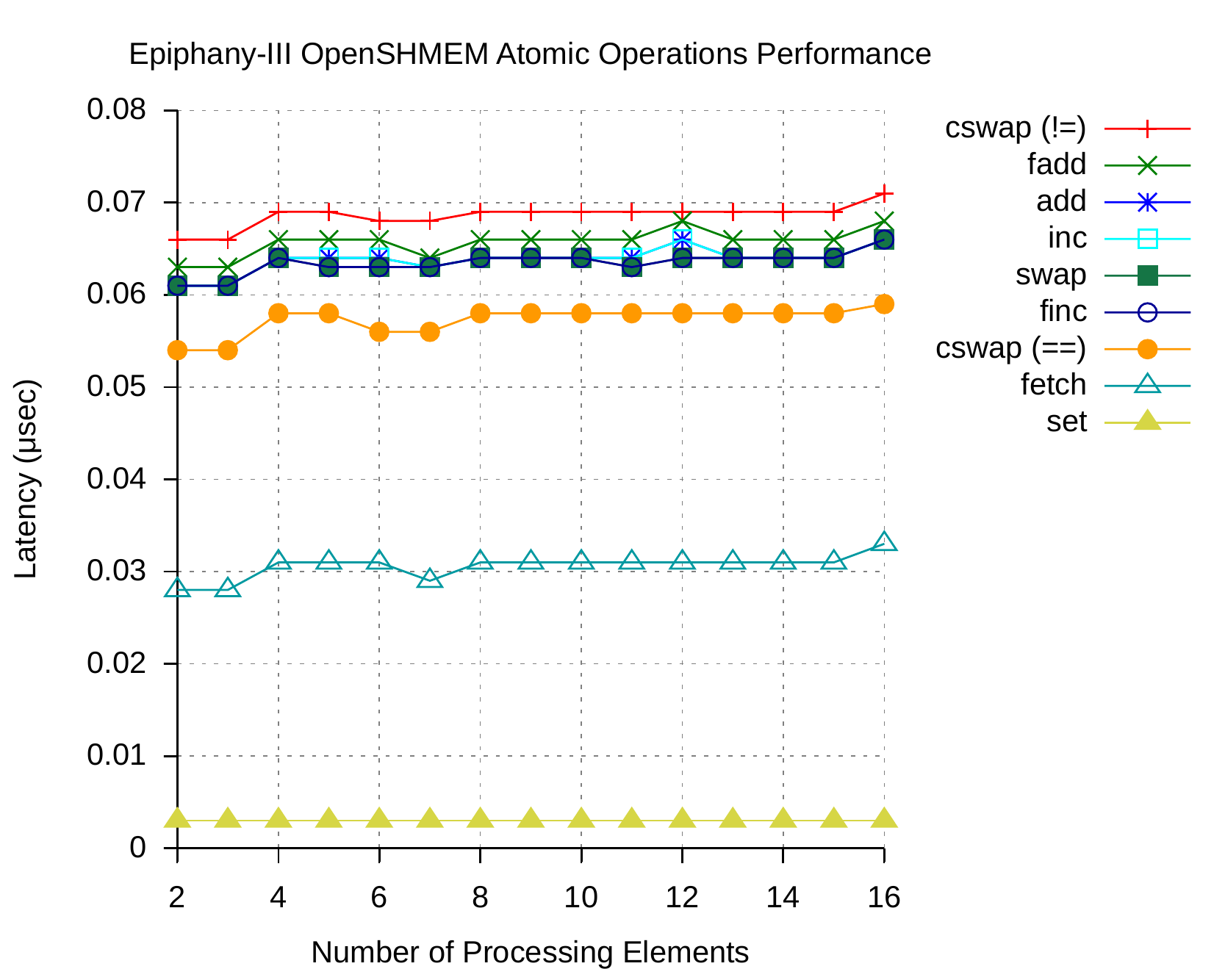}
	\caption{Performance of \mbox{OpenSHMEM} atomic operations for 32-bit
integers and a variable number of processing elements. Atomic operations are
performed in a tight loop on the next neighboring processing element.}
	\label{fig:atomics}
\end{figure}

The Epiphany-III ISA does not have support for atomic instructions, but the
\texttt{TESTSET} instruction used for remote locks may be used to define other
atomic operations in software. With the current code design, it is trivial to
extend to other atomic operations with a single line of code if additional
atomic operations are defined by the \mbox{OpenSHMEM} specification in the
future. At a core level, memory access for both fetch and set operations
completes in a single clock cycle and is therefore implicitly atomic. The
fetch operation still must traverse the network to the remote core and return
the result. Each data type specialization uses a different lock on the remote
core as per the specification. The performance results for the 32-bit integer
atomic routines appear in Figure~\ref{fig:atomics}.

\subsection{Collective Routines}

\begin{figure}
	\centering
		\includegraphics[width=0.49\textwidth]{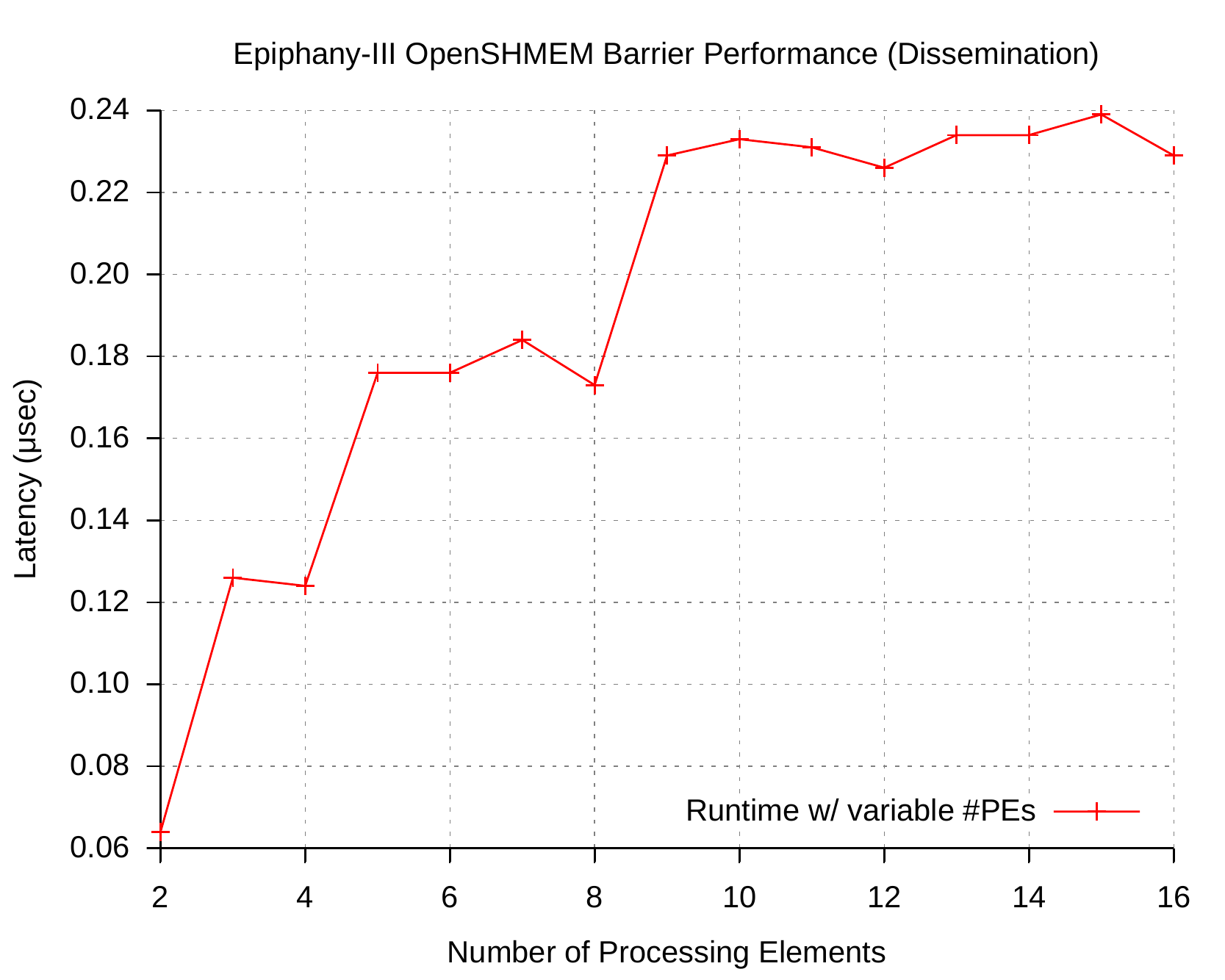}
		\includegraphics[width=0.49\textwidth]{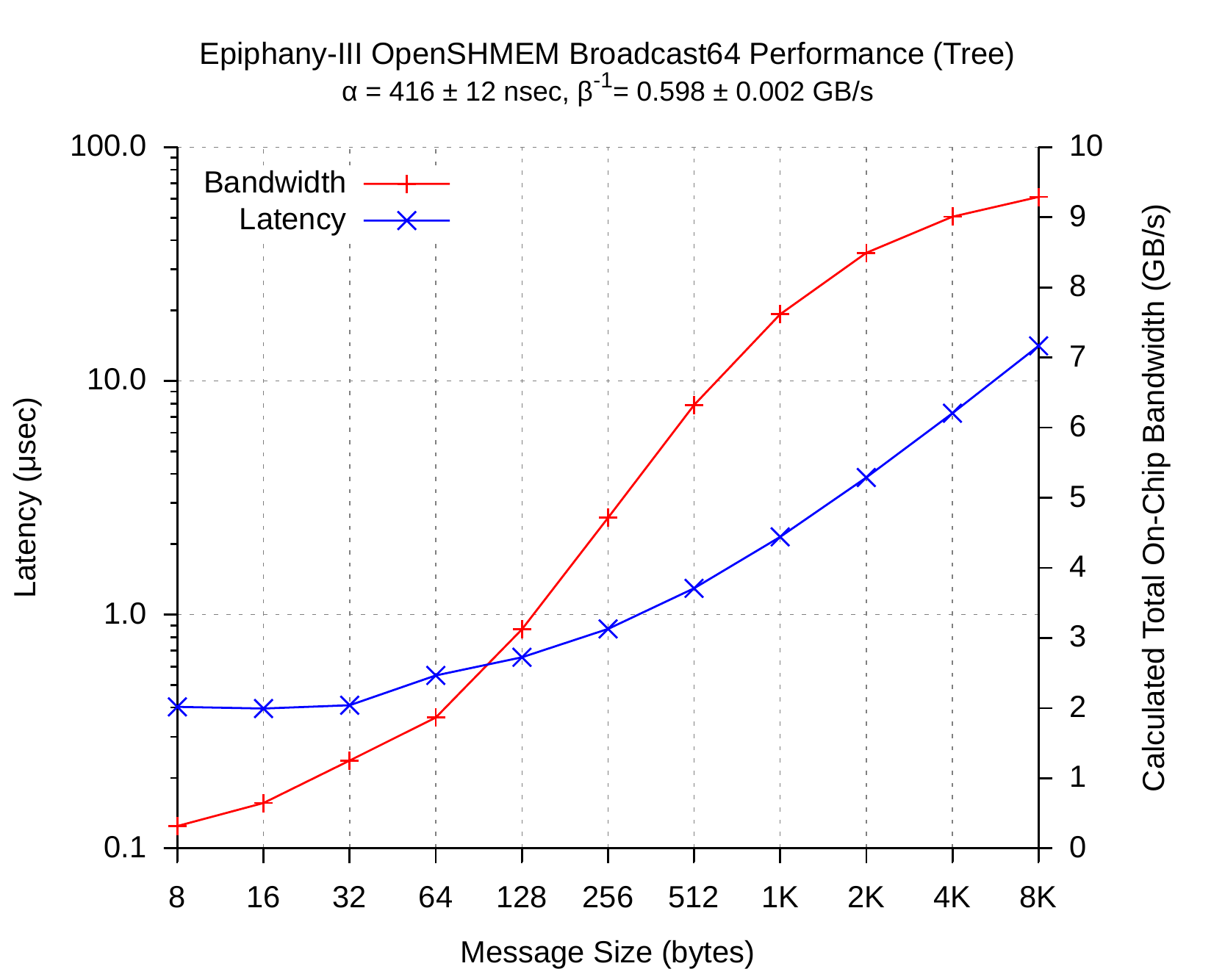}
	\caption{Performance of \texttt{shmem\_barrier} for variable number of
processing elements (left) and the performance of \texttt{shmem\_broadcast64}
for variable message sizes (right)}
	\label{fig:barrier_broadcast}
\end{figure}

Multi-core barriers are critical to performance for many parallel applications.
The Epiphany-III includes hardware support for a fully collective barrier with
the \texttt{WAND} instruction and corresponding ISR. This hardware support is
included as an experimental feature within the \mbox{OpenSHMEM} library and
must be enabled by specifying \texttt{SHMEM\_USE\_WAND\_BARRIER} at compile
time. After several implementations of barrier algorithms, it was determined
that a dissemination barrier was the highest-performing software barrier
method. It is not clear if this algorithm will continue to achieve the
highest performance on chip designs with a larger number of cores;
alternative tree algorithms may be needed. The eLib interface in the eSDK uses
a counter-based collective barrier and requires a linearly increasing amount of
memory with the number of cores. The dissemination barrier requires \(8 \cdot
log_2(N)\) bytes of memory, where \emph{N} is the number of processing elements
within the barrier. The use of this synchronization array mitigates the need
for signaling by locks at each stage of the barrier. The collective eLib
barrier completes in 2.0~$\mu$sec while the WAND barrier completes in
0.1~$\mu$sec. The performance for group barriers for a subset of the total
processing elements is shown in Figure~\ref{fig:barrier_broadcast}. The latency
of the dissemination barrier increases logarithmically with the number of cores
so that more than eight cores take approximately 0.23~$\mu$sec.


Broadcasts are important in the context of the Epiphany application development
in order to limit the replication of off-chip memory accesses to common memory.
It is faster to retrieve off-chip data once and disseminate it to other
processing elements in an algorithmic manner than for each processing element
to fetch the same off-chip data. The data are distributed with a logical
network tree, moving the data the farthest distance first in order to prevent
subsequent stages increasing on-chip network congestion. The broadcast routines
use the same high-performance memory copying subroutine as the contiguous data
transfers. Effective core bandwidth approaches the theoretical peak performance
for this algorithm and is approximately \(2.4 / log_2(N)\) GB/sec.
Figure~\ref{fig:barrier_broadcast} shows collective broadcast performance for
variable message sizes.

\begin{figure}
	\centering
		\includegraphics[width=0.49\textwidth]{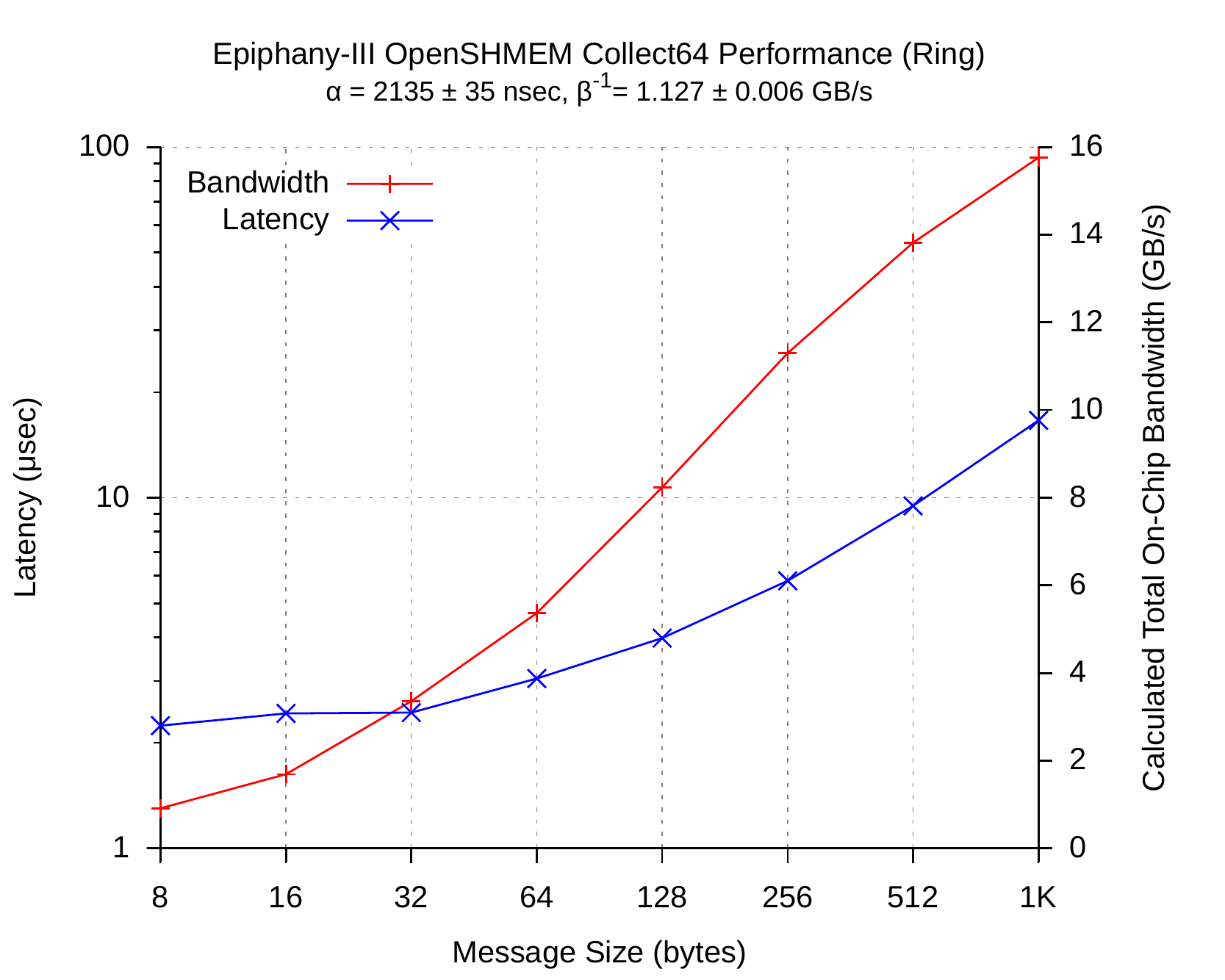}
		\includegraphics[width=0.49\textwidth]{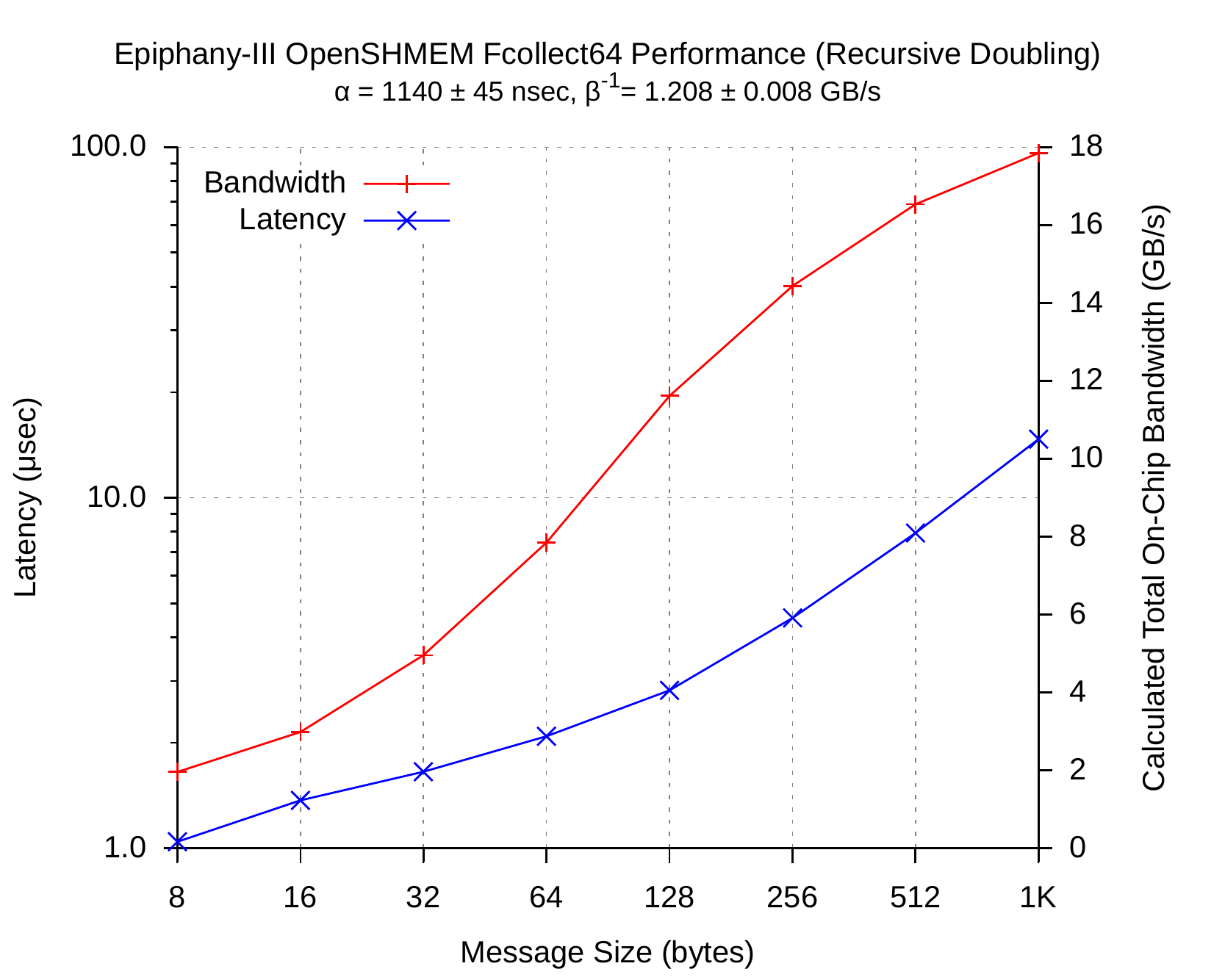}
	\caption{Performance of linear scaling \texttt{shmem\_collect64} and recursive
doubling \texttt{shmem\_fcollect64} for variable message sizes on 16~processing
elements}
	\label{fig:collect}
\end{figure}

The \texttt{shmem\_collect} and \texttt{shmem\_fcollect} routines use ring and
recursive doubling algorithms for concatenating blocks of data from multiple
processing elements. Each uses the optimized contiguous memory copying routine.
There is likely room for improvement with these routines; the measured
performance appears in Figure~\ref{fig:collect}.

\begin{figure}
	\centering
		\includegraphics[width=0.67\textwidth]{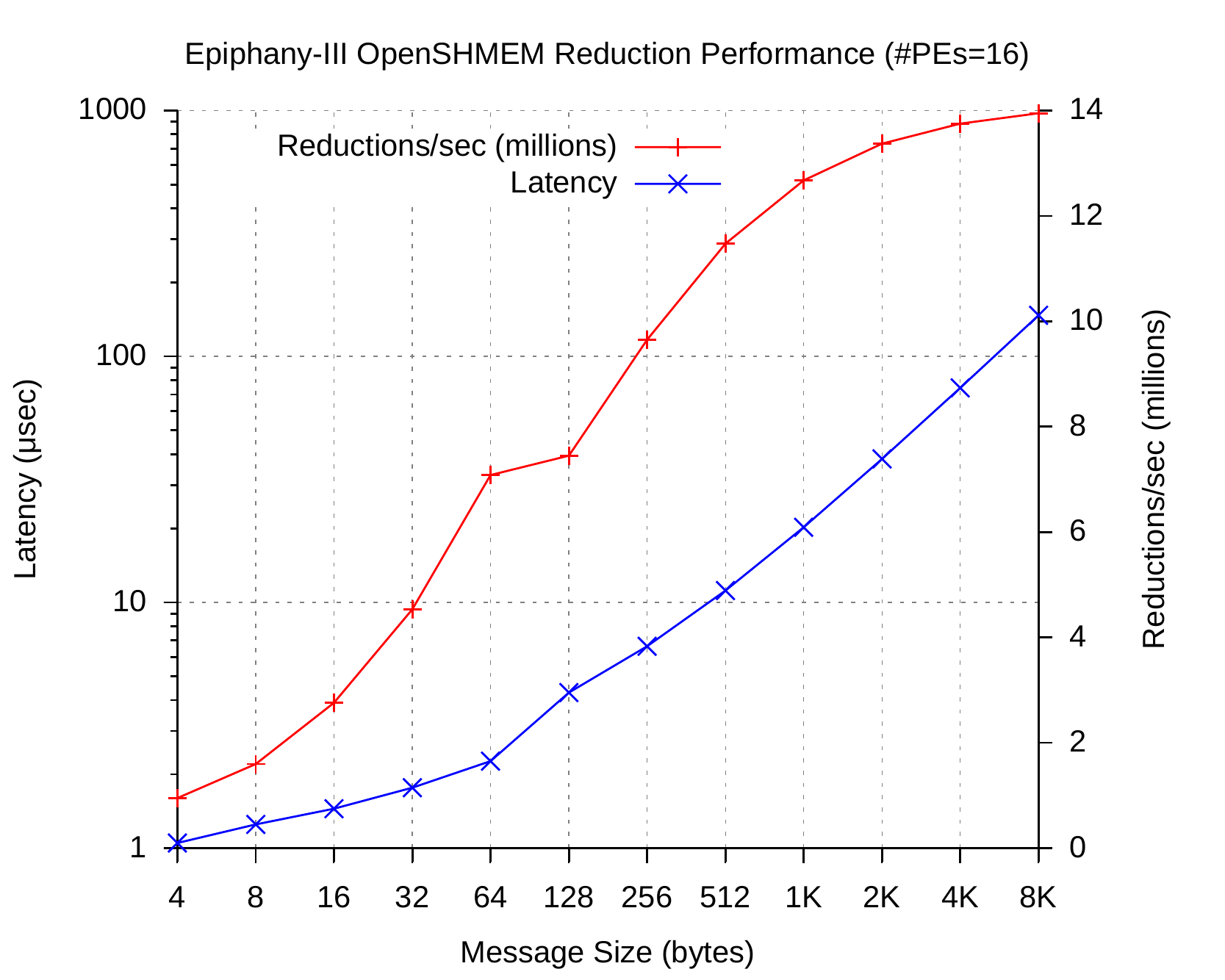}
	\caption{Reduction performance for \texttt{shmem\_int\_sum\_to\_all} for all
16~processing elements. The latency and the number of collective reductions per
second are shown. The effect of the minimum symmetric work array size for
reductions, defined as \texttt{SHMEM\_REDUCE\_MIN\_WRKDATA\_SIZE} per the
\mbox{OpenSHMEM} specification, is apparent for small reductions}
	\label{fig:reduction}
\end{figure}

The \texttt{shmem\_\emph{TYPE}\_\emph{OP}\_to\_all} reduction routines are
important for many multi-core applications. The routines use different
algorithms depending on the number of processing elements. A ring algorithm is
used for processing elements that number in non-powers of two and a
dissemination algorithm for powers of two. The symmetric work array is used for
temporary storage and the symmetric synchronization array is used for
multi-core locks and signaling. The performance of
\texttt{shmem\_int\_sum\_to\_all} appears in Figure~\ref{fig:reduction}. Other
routines vary marginally in performance due to data types and the arithmetic
operation used. Reductions that fit within the symmetric work array have
improved latency as seen in the figure.

\begin{figure}
	\centering
		\includegraphics[width=0.67\textwidth]{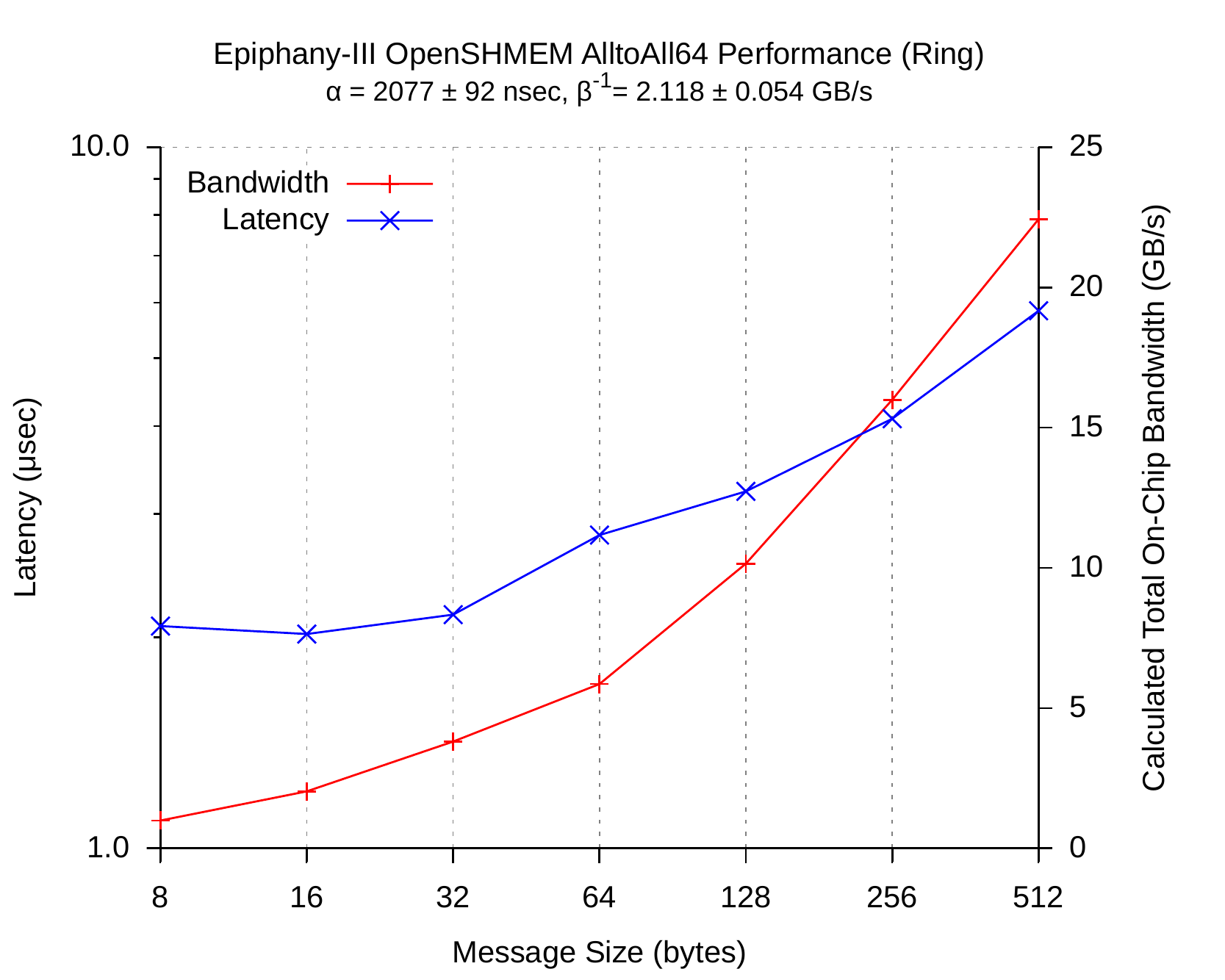}
	\caption{Performance of the new (to version~1.3) contiguous all-to-all data
exchange operation, \texttt{shmem\_alltoall}, for 16~processing elements}
	\label{fig:alltoall}
\end{figure}

The performance of the contiguous all-to-all data exchange,
\texttt{shmem\_alltoall}, appears in Figure~\ref{fig:alltoall}. This routine
has a relatively high overhead latency compared to other collectives.

\subsection{Distributed Locking Routines}

The distributed locking routines, \texttt{shmem\_set\_lock} and
\texttt{shmem\_test\_lock}, are easily supported by the atomic \texttt{TESTSET}
instruction. The actual lock address is defined in the implementation to be on
the first processing element. These locking mechanisms are also the basis for
the atomic operations detailed in Sect. \ref{ssec:atomic} but for multiple
processing elements. The \texttt{shmem\_clear\_lock} routine is a simple remote
write to free the lock. Although this scheme works well for the 16 processing
elements on the Epiphany-III, the performance bottleneck will likely be a
problem scaling to much larger core counts. Application developers should avoid
using these global locks.

\section{Future Work and Discussion of Extensions for Embedded Architectures}
\label{sec:future}

It is our intention to release \emph{ARL \mbox{OpenSHMEM} for Epiphany}, as
well as the performance evaluation codes and benchmarks used in this paper, as
open source software through the U.S. Army Research Laboratory's GitHub account
\cite{arlgithub} for Parallella community input and further development. The
Epiphany architecture may also be updated in the future to add more hardware
support for many of the existing \mbox{OpenSHMEM} routines. Many of the
currently proposed \mbox{OpenSHMEM} extensions and updates should be
addressable. A non-blocking strided remote memory access routine could be
supported with the existing DMA engine as mentioned in Sect.
\ref{ssec:nonblocking}. Some other extensions do not make sense for the
architecture. For example, Epiphany is not a multithreaded architecture and,
although it can be performed via software, is not the ideal mechanism for
improving performance. The \mbox{OpenSHMEM} standard should remain sufficiently
lightweight to address low-level operations without relying on specific
architectural features.

One of the more challenging portions of the \mbox{OpenSHMEM} standard for the
Epiphany architecture and other embedded architectures are the memory
management routines. It makes some sense for some platforms to have a
pre-allocated symmetric heap from which memory allocations will be made. Within
an Epiphany local core, there is no memory virtualization between the physical
address and the memory address returned by the allocation routines as available
memory is linearly removed from the symmetric heap. The limitations of the
available core space make it challenging to introduce a Linux-like abstract
model of virtual memory. As \mbox{OpenSHMEM} is a low-level interface and
application developers are already accustomed to explicitly managing memory, it
may make some sense to improve memory management interfaces, such as those
discussed in Sect. \ref{ssec:memory}, for embedded architectures.

\section{Conclusion}
\label{sec:conclusion}

\mbox{OpenSHMEM} provided an effective and pragmatic programming model for the
Epiphany architecture. The header-only implementation enabled compiler
optimizations for program size and application performance that is difficult to
achieve using a standard pre-compiled library. We demonstrated improved
performance and many useful features compared to the current eLib library
despite the additional software abstraction with the OpenSHMEM interface. The
\emph{ARL \mbox{OpenSHMEM} for Epiphany} demonstrated high-performance
execution while approaching hardware theoretical networking limits, and
low-latency operation for many of the \mbox{OpenSHMEM} routines.

\bibliography{OpenSHMEM2016}

\end{document}